\documentclass[12pt]{article}
\usepackage[english,german,french,polish]{babel}
\usepackage{amsfonts}

\begin{document}

\selectlanguage{english}

\baselineskip 0.73cm
\topmargin -0.4in
\oddsidemargin -0.1in

\let\ni=\noindent

\renewcommand{\thefootnote}{\fnsymbol{footnote}}

\newcommand{\SM}{Standard Model }

\pagestyle {plain}

\setcounter{page}{1}



~~~~~~
\pagestyle{empty}

\begin{flushright}
IFT-- 09/4
\end{flushright}

\vspace{0.4cm}

{\large\centerline{\bf More about the hypothesis of a new weak interaction}}

{\large\centerline{\bf of electromagnetic field in the hidden sector}}

\vspace{0.5cm}

{\centerline {\sc Wojciech Kr\'{o}likowski}}

\vspace{0.3cm}

{\centerline {\it Institute of Theoretical Physics, University of Warsaw}}

{\centerline {\it Ho\.{z}a 69, 00--681 Warszawa, ~Poland}}

\vspace{0.6cm}

{\centerline{\bf Abstract}}

\vspace{0.2cm}

\begin{small}


New hypothetical field equations (Eqs. (1) and (2)) are further discussed, unifying Maxwell's equations of the 
Standard Model (after the electroweak symmetry is spontaneously broken) with the dynamics of hidden sector 
(expected to be responsible for the cold dark matter). The hidden sector is represented by sterile spin-1/2 Dirac 
fermions ("sterinos") and sterile spin-0 bosons ("sterons") whose masses are spontaneously generated by a nonzero
vacuum expectation value of the steron field, while sterino and steron interactions are mediated by sterile spin-1 
quanta of an antisymmetric-tensor field with a large mass scale ("$A$ bosons"). These interactions are presumed to be 
weak, but stronger than the universal gravity. Beside sterinos and sterons, the Standard-Model photons are 
included into the source of sterile $A$ bosons and so, they become a link between the hidden and 
Standard-Model sectors ("photonic portal"\, to the hidden sector). The relativistic structure of antisymmetric-tensor field 
of sterile $A$ bosons  can be split into a vector and an axial three-dimensional fields (of spin 1 and parities -- and +) 
in such a way that the Standard-Model electric and magnetic fields become involved separately in the sources of these two kinds of sterile $A$-boson radiation, respectively. 
 
\vspace{0.6cm}

\ni PACS numbers: 14.80.-j , 04.50.+h , 95.35.+d 

\vspace{0.3cm}

\ni April 2009

 
\end{small}

\vfill\eject

\pagestyle {plain}

\setcounter{page}{1}

\vspace{0.5cm}

\ni {\bf 1. Sterile mediating field in the hidden sector and the photon portal}

\vspace{0.4cm} 

In a recent work [1], we considered the hypothesis that, beside the Standard-Model sector of the Universe, there is a hidden sector consisting of massive spin-1/2 and spin-0 sterile particles (called here "sterinos"\, and "sterons", respectively). They interact weakly through massive spin-1 sterile particles (called "$\!A\!$ bosons") which collaborate with the Standard-Model photons participating jointly with sterinos and sterons in the source of {\it A} bosons, 

\begin{equation}
(\Box - M^2)A_{\mu\,\nu} = - \sqrt{f} (\varphi F_{\mu\,\nu} + \zeta \bar\psi \sigma_{\mu\,\nu} \psi)\;, 
\end{equation}

\ni where {\it A} bosons are described by an antisymmetric-tensor field $A_{\mu\,\nu}$ (of dimension one) and $F_{\mu\,\nu} = \partial_\mu A_\nu - \partial_\nu A_\mu$ is the Standard-Model electromagnetic field (of dimension two), while $\psi$ and $\varphi$ stand for sterino and steron fields. In Eq. (1), $f$ and $\zeta$ denote two dimensionless constants and $M$ is a large mass scale.

Simultaneously, in the presence of hidden sector, the Maxwell's equations for Standard-Model electromagnetic field $F_{\mu \nu}$ become supplemented to the form

\begin{equation}
\partial^\nu (F_{\mu\,\nu} +  \sqrt{f} \varphi A_{\mu\,\nu}) = -j_\mu \;,\; F_{\mu\,\nu} = \partial_\mu A_\nu - \partial_\nu A_\mu 
\end{equation}

\ni with  $j_\mu$ still being the Standard-Model electric current. In Eqs. (1) and (2) the expansion

\begin{equation} 
\varphi = <\!\!\varphi\!\!>_{\rm vac} + \varphi^{\rm (ph)}
\end{equation}

\ni holds, where $<\!\!\varphi\!\!>_{\rm vac}$ and $\varphi^{\rm (ph)}$ are a spontaneously-nonzero vacuum expectation value of the field $\varphi$ and the physical steron field, respectively.

Of course, the field equations (1) and (2) can be valid when the Standard-Model electroweak symmetry is spontaneously broken by the Standard-Model Higgs mechanism and photons emerge. The new weak interaction of sterinos and sterons mediated by $A$ bosons collaborating with Standard-Model photons is presumed to be stronger than the very weak interaction provided by universal gravity. In a natural way, the hidden sector is expected to be responsible for the cold dark matter, in particular, for its thermal relic abundance [2,3].

At the same time, when Eqs. (1) and (2) hold, the sterino and physical-steron fields $\psi$ and $\varphi^{\rm (ph)}$  satisfy the field equations (being sterile\,"matter equations")

\begin{equation}
\left(i \gamma^\mu \partial_\mu - \frac{1}{2}\sqrt{f}\,\zeta \sigma_{\mu\,\nu} A^{\mu\,\nu} - m_\psi\right) \psi = 0 
\end{equation}

\ni and

\begin{equation}
(\Box - m^2_\varphi) \varphi^{\rm (ph)} = \frac{1}{2}\sqrt{f} \, F_{\mu\,\nu} A^{\mu\,\nu} \;, 
\end{equation}

\ni where $m_\psi$ and $m_\varphi$ are sterino and physical-steron masses.

All four kinds of field equations for $A_{\mu \nu}$, $F_{\mu \nu}$ and $\psi$, $\varphi^{\rm (ph)}$ are mutually consistent, as they can be derived from the following Lagrangian which is a supplemented form of the Standard-Model Lagrangian  $ - (1/4) F_{\mu \nu} F^{\mu \nu} - j_\mu A^\mu $ in the presence of hidden sector:  

\begin{eqnarray}
{\cal{ L }} & = &  -\frac{1}{4}F_{\mu\,\nu} F^{\mu\,\nu} - j_\mu A^\mu
- \frac{1}{2} \sqrt{f}\left(\varphi F_{\mu \nu} + \zeta \bar\psi \sigma_{\mu\,\nu} \psi \right) A^{\mu \nu}
\nonumber \\ & &   - \frac{1}{4} \left[ \left(\partial_\lambda A_{\mu \nu}\right)\left(\partial^\lambda A^{\mu \nu}\right)  
- M^2 A_{\mu \nu}A^{\mu \nu}\right]  \nonumber \\
 & & + \,\bar\psi \left(i \gamma^\lambda \partial_\lambda - m_\psi  \right) \psi + \left[(\partial_\lambda\varphi^{\rm (ph)} ) (\partial^\lambda \varphi^{\rm (ph)}  ) - m^2_\varphi \varphi^{{\rm (ph)}\,2} \right]  
\end{eqnarray}

\ni with $F_{\mu\,\nu} = \partial_\mu A_\nu - \partial_\nu A_\mu $ and $\varphi = <\!\!\varphi\!\!>_{\rm vac} + \varphi^{\rm (ph)}$. Of course, the total Lagrangian (corresponding to $\cal{L}$ given in Eq. (6)) includes also the 
Standard-Model Lagrangian [4] minus its electromagnetic part $ - (1/4) F_{\mu \nu} F^{\mu \nu} - j_\mu A^\mu $ which in Eq. (6) is shifted to the hidden-sector Lagrangian. Here, the masses of sterile particles can be considered as spontaneously generated by $ <\!\!\varphi\!\!>_{\rm vac} \neq 0$: $M = \sqrt{ \eta/2} <\!\!\varphi\!\!>_{\rm vac}$, $m_\psi = \xi <\!\!\varphi\!\!>_{\rm vac}$ and  $m_\varphi = \sqrt{ \lambda/2}\, {<\!\!\varphi\!\!>_{\rm vac}}$, where $\eta, \xi$ and $\lambda$ denote dimensionless constants appearing in the hidden-sector Lagrangian before the spontaneous mass generation is carried out for sterile particles ({\it cf.} the last Ref. [1], in particular, Eqs. (21), (14) and (13)). 

Due to the new additional weak interaction of photons{\footnote{The phenomenon of different interactions of the same particles is typical in particle physics, leading usually to the interaction unification. For instance, leptons and quarks display, in addition to the universal gravity, two and three phenomenologically different interactions, respectively, unified into the spontaneously broken electroweak symmetry of \SM and, possibly, symmetries of GUTs. Our hypothetical weak interaction of photons in the hidden sector (after the electroweak symmetry is spontaneously broken by the Standard-Model Higgs mechanism and photons emerge) might be unified with their original electromagnetic interaction in the Standard-Model sector along the line of sterino magnetic moment spontaneously generated by $<\!\!\varphi\!\!>_{\rm vac} \neq 0$ ({\it cf.} Eq. (10)).}} described by the term $- \frac{1}{2} \sqrt{f} \varphi F_{\mu \nu} A^{\mu \nu}$ in the Lagrangian (6) (where their original electromagnetic interaction is given by the term $-j_\mu A^\mu $ with $j_\mu$ being the Standard-Model electric current), the hidden sector can interact weakly with the Standard-Model sector  {\it via} photons interacting in both sectors (this mechanism is called "photonic portal").

If $-\Box$ can be neglected in Eq. (1) versus $M^2$ (in an analogy to the familiar case of Fermi interaction), we can write approximately 

\begin{equation}
 A_{\mu \nu} \simeq 
\frac{\sqrt{f}}{M^2}\left(\varphi F_{\mu \nu} + \zeta \bar\psi \sigma_{\mu\,\nu} \psi \right)\,.
\end{equation}

\ni Then, eliminating $A_{\mu \nu}$ from the interaction term in the Lagrangian (6) and dividing the results by 2 to avoid double counting in such an operation leading to a quadratic form in the effective Lagrangian, we obtain approximately the following effective interaction (being our sterile analogy of Fermi interaction):

\begin{equation}
  -\frac{1}{4} \frac{f}{M^2}\left(\varphi F_{\mu \nu} + \zeta \bar\psi \sigma_{\mu\,\nu} \psi \right) \left(\varphi F^{\mu \nu} + \zeta \bar\psi \sigma^{\mu\,\nu} \psi \right) \,.
\end{equation}

\ni In the absence of physical sterons (when $\varphi = <\!\!\varphi\!\!>_{\rm vac}$), the cross term in the effective interaction (8) gives

\begin{equation}
- \frac{f \zeta <\!\!\varphi\!\!>_{\rm vac}}{2M^2} \bar\psi \sigma_{\mu\,\nu} \psi F^{\mu \nu}  \equiv -\mu_\psi \bar\psi \sigma_{\mu\,\nu} \psi F^{\mu \nu}\,,  
\end{equation}

\ni showing that here a weak effective sterino magnetic moment

\vspace{-0.2cm}

\begin{equation}
 \mu_\psi = \frac{f \zeta}{2M^2}<\!\!\varphi\!\!>_{\rm vac} 
\end{equation}

\ni is spontaneously generated by $<\!\!\varphi\!\!>_{\rm vac} \neq 0$. Of course, neutral sterinos get here no effective electric charge: $Q_\psi = 0$ (so, the equality $\mu_\psi = Q_\psi/2m_\psi$ does not work). In fact, the correction to the electric current $j_\mu$ in Eq. (2),

\begin{equation}
\delta j_\mu = \partial^\nu(\sqrt{f} \varphi A_{\mu \nu}) \,,
\end{equation}

\ni gives a zero correction to the electric charge $Q = \int d^3x j_0$:

\begin{equation}
\delta Q = \int d^3x \delta j_0 = \int d^3x \partial^l(\sqrt{f} \varphi A_{0 l} ) = 0\,.
\end{equation}

We can see that at least a part of cold dark matter expected to consist of sterinos (whose mass and magnetic moment are spontaneously generated by $<\!\!\varphi\!\!>_{\rm vac} \neq 0$) is magnetic in the sense that it can interact weakly with cosmic and laboratory magnetic fields. In particular, it may be weakly polarized in external magnetic fields.

\vspace{0.4cm}

\ni {\bf 2. Radiation of sterile{\it A} bosons} 

\vspace{0.4cm}

The antisymmetric-tensor field $A_{\mu \nu}$ of mediating sterile {\it A} bosons can be expressed by a vector and an axial 
three-dimensional fields $\vec{A}^{\rm (E)}$ and $\vec{A}^{\rm (B)}$ of spin 1 and parities -- and +, respectively, defined as follows:

\begin{equation}
A_{k 0} = - A^{\rm (E)}_k \; \;,\; \; A_{k l} = - \varepsilon_{k l m} A^{\rm (B)}_m \; \;,\; \; A_{0 l} =  A^{\rm (E)}_l \;, 
\end{equation}

\ni where $\vec{A}^{\rm (E)} = \left(A^{\rm (E)}_k \right)$ and $\vec{A}^{\rm (B)} = \left(A^{\rm (B)}_k \right)\;(k,l = 1,2,3)$. Note that for the electromagnetic field $F_{\mu\,\nu} = \partial_\mu A_\nu - \partial_\nu A_\mu$  we write $(A^\mu) =  (A^0,\vec{A})$ with $A^0 = A_0$ and $\vec{A} = (A^k) = (-A_k)$, and $(\partial_\mu) = (\partial/\partial{x}^\mu) = (\partial_0,\vec{\partial})$ with $\partial_0 = \partial^0$ and $\vec{\partial} = (\partial_k) = (-\partial^k)$, whereas $(x^\mu) = (x^0,\vec{x})$ with $x^0 = x_0$ and $\vec{x} = (x^k) = (-x_k)$. From Eq. (13) it follows that $A^{\rm (B)}_k = -(1/2)\varepsilon_{k l m} A_{l m}$. Here, $\varepsilon_{123} = \varepsilon^{123} = 1$ though $-\varepsilon_{0123} = \varepsilon^{0123} = 1$.

From the definitions (13) we obtain

\begin{equation} 
\left(A_{\mu \nu}\right) = \left(\begin{array}{rrrr} 0\;\;\;\;  & A^{\rm (E)}_1 & A^{\rm (E)}_2 & A^{\rm (E)}_3 \\ -A^{\rm (E)}_1 & 0\;\; \;\; & -A^{\rm (B)}_3 & A^{\rm (B)}_2 \\ -A^{\rm (E)}_2 & A^{\rm (B)}_3 & 0\;\; \;\; & -A^{\rm (B)}_1 \\  -A^{\rm (E)}_3 & -A^{\rm (B)}_2 & A^{\rm (B)}_1 & 0\;\; \;\; \end{array} \right)  = \left(A^{\rm (E)}_{\mu \nu}\right) + \left(A^{\rm (B)}_{\mu \nu}\right) 
\end{equation}

\ni with the separated $A^{\rm (E)}_{\mu \nu}$ and $A^{\rm (B)}_{\mu \nu}$. On the other hand, for the electromagnetic field we have

\begin{equation} 
\left(F^{\mu \nu}\right) = \left(\begin{array}{rrrr} 0\;\;  & -E_1 & -E_2 & -E_3 \\ E_1 & 0\;\; & -B_3 & B_2 \\ E_2 & B_3 & 0\;\; & -B_1 \\  E_3 & -B_2 & B_1 & 0\;\; \end{array} \right) \,,
\end{equation}

\ni since 

\begin{equation} 
\vec{E} = -\partial_0 \vec{A} - \vec{\partial} A_0 = (\partial_0 A_k - \partial_k A_0) =(-F_{k  0})\;,\; \vec{B} = 
\vec{\partial}\times \vec{A} = (-\varepsilon_{klm}\partial_l A_m)= (-\frac{1}{2}\varepsilon_{klm} F_{l m})\,,
\end{equation}

\ni where $\vec{E} = (E_k)$ and $\vec{B} = (B_k)$, while $\partial_0 = \partial/\partial x^0$ and $\vec{\partial} = \partial/\partial \vec{x} = (\partial_k)$. From Eqs. (14) and (15) it follows that 

\begin{equation} 
F^{\mu \nu} A_{\mu \nu} = - 2\left(\vec{E}\cdot \vec{A}^{\rm (E)} - \vec{B}\cdot \vec{A}^{\rm (B)}\right) 
\end{equation}

\ni and

\begin{equation} 
F^{\mu \nu} F_{\mu \nu} = - 2\left(\vec{E}^2 - \vec{B}^2\right)\;\;\;,\;\;\;A^{\mu \nu} A_{\mu \nu} =  -2\left( \vec{A}^{\rm (E)\,2} - \vec{A}^{\rm (B)\,2}\right) \,.
\end{equation}

For the Dirac spin tensor $\sigma^{\mu \nu} = (i/2)[\gamma^\mu,\gamma^\nu]$ giving explicitly

\begin{equation} 
(\sigma^{\mu \nu}) = \left(\begin{array}{rrrr} 0\;\;  &i \alpha_1 & i \alpha_2 & i \alpha_3 \\ -i \alpha_1 & 0\;\; & \sigma_3 & -\sigma_2 \\ -i \alpha_2 & -\sigma_3 & 0\;\; & \sigma_1 \\  -i \alpha_3 & \sigma_2 & -\sigma_1 & 0\;\; \end{array} \right) 
\end{equation}

\ni with $\vec{\alpha} = (\alpha_k) = (\gamma^0\gamma^k)$ and $\vec{\sigma}=(\sigma_k) = \gamma_5\vec{\alpha} = (\gamma_5 \gamma^0\gamma^k)$, where $\sigma_{kl} = \varepsilon_{k l m} \sigma_m$ and $\sigma_k = (1/2)\varepsilon_{k l m} \sigma_{l m}$, we get

\begin{equation} 
\sigma^{\mu \nu}A_{\mu \nu} = 2\left(i \vec{\alpha}\cdot \vec{A}^{\rm (E)} - \vec{\sigma}\cdot \vec{A}^{\rm (B)}\right)\,.
\end{equation}

Due to Eqs. (17) and (20), the interaction term in the Lagrangian (6) can be rewritten in the form

\begin{equation}
-\frac{1}{2}\sqrt{f}\left(\varphi F^{\mu \nu} + \zeta \bar\psi \sigma^{\mu\,\nu} \psi \right)\!A_{\mu \nu} = \sqrt{f}\left[\left(\varphi \vec{E} - i \zeta \bar{\psi} \,\vec{\alpha}\,\psi \right)\!\cdot\!\vec{A}^{\rm (E)} - \left(\varphi \vec{B} - \bar{\psi}\,\vec{\sigma}\,\psi \right)\!\cdot\! \vec{A}^{\rm (B)}\right]\,.
\end{equation}

In a similar way, the field equation (1) for $A_{\mu \nu}$ can be presented in the three-dimensional form

\begin{eqnarray} 
(\Box - M^2)\vec{A}^{\rm (E)} & = & - \sqrt{f}\left(\varphi \vec{E} - i \zeta \bar{\psi}\,\vec{\alpha}\,\psi \right)\, \nonumber \\ (\Box - M^2)\vec{A}^{\rm (B)} & = & - \sqrt{f}\left(\varphi \vec{B}\, - \,\zeta\bar{\psi}\, \vec{\sigma}\,\psi \right)\,.
\end{eqnarray}

\ni Here, as usual, $\varphi = <\!\!\varphi\!\!>_{\rm vac} + \varphi^{\rm (ph)}$ with $<\!\!\varphi_{\rm vac}\!\!> \neq 0$.

We can conclude from Eqs. (21) or (22) that the electric and magnetic fields $\vec{E}$ and $\vec{B}$, parts of $F_{\mu \nu}$, are involved respectively in the sources of two different sterile fields $\vec{A}^{\rm (E)}$ and $\vec{A}^{\rm (B)}$, parts of the relativistic structure $A_{\mu \nu}$. In particular, in the absence of physical sterons (when $\varphi = <\!\!\varphi\!\!>_{\rm vac}$), these sources are given effectively by
 
\begin{equation} 
-\sqrt{f}\left(<\!\!\varphi\!\!>_{\rm vac}\vec{E} - i \zeta \bar{\psi}\,\vec{\alpha}\,\psi \right)\;\; \,,\,\;\; -\sqrt{f}\left(<\!\!\varphi\!\!>_{\rm vac}\vec{B}\, - \,\zeta \bar{\psi}\, \vec{\sigma}\,\psi \right)\,.
\end{equation}

\ni If there are also no sterinos (when $\psi = 0$), these sources are reduced effectively to

\begin{equation} 
-\sqrt{f}<\!\!\varphi\!\!>_{\rm vac}\vec{E} \;\; \,,\,\;\; -\sqrt{f}<\!\!\varphi\!\!>_{\rm vac}\vec{B}\,.
\end{equation}

The sterile $A$ bosons of two kinds described by the fields $\vec{A}^{\rm (E)}$ and $\vec{A}^{\rm (B)}$, when they participate in a sterile radiation propagating freely in space, get the following wave functions: 

\begin{equation} 
\vec{A}^{\,\rm (E,B)}_{\vec{k}}(x) = \frac{1}{(2\pi)^{3/2}} \frac{1}{\sqrt{2 \omega_A}}\, \vec{e}^{\rm \,(E,B)} e^{-i k_A\cdot x } \,,
\end{equation}

\ni where $k_A = (k^\mu_A) = (\omega_A, \vec{k}_A)$ with $\omega_A =\sqrt{\vec{k}_A^2 + M^2}$, while $\vec{e}^{\rm \,(E,B)} = \vec{e}_a^{\rm \,(E,B)}\;\;(a=1,2,3)$ are three orthonormal linear polarizations for $(E)$ and $(B)$ kinds of $A$ bosons, $A^{\rm (E)}$ and $A^{\rm (B)}$, satisfying the formulae

\begin{equation} 
\vec{e}_a^{\rm \,(E,B)}\cdot \vec{e}_b^{\rm \,(E,B)} = \delta_{a b}\;\; (a,b = 1,2,3) \;\;,\;\; \sum^3_{a=1} {e}_{a k}^{\rm \,(E,B)} {e}_{a l}^{\rm \,(E,B)} = \delta_{k l}
\end{equation}

\ni with $\vec{e}_a ^{\rm \,(E,B)} = ({e}_{a k}^{\rm \,(E,B)})$.

The annihilation operators $a_{\vec{k}}^{\rm \,(E,B)}$ standing at the wave functions (25) in the Fourier expansions of quantum fields $\vec{A}^{\rm (E,B)}(x)$ annihilate respectively vector or axial sterile $A$ bosons of spin 1 and parity -- or +, propagating in the wave states $\vec{A}^{\rm (E,B)}_{\vec{k}}(x)$.

In the next two Sections we will calculate in the lowest order the decay rates of free sterile $A$ bosons of two kinds into free particle pairs $\varphi^{(\rm ph)} \gamma$ and $e^+e^-$. On the other hand, the free sterile $A$ bosons can be produced in, for instance, the inelastic Compton scattering $e^-\gamma \rightarrow e^-\gamma* \rightarrow e^- A$ and $p\gamma \rightarrow p \gamma^* \rightarrow p A$ with $\gamma^* \rightarrow A$ (when $\varphi = <\!\!\varphi\!\!>_{\rm vac}$). 

\vspace{0.4cm}

\ni {\bf 3. Decays of $A^{\,\rm (E,B)} \rightarrow \varphi^{(\rm ph)} \gamma $ (with $M>m_\varphi$)}

\vspace{0.4cm}

Consider two processes $A^{\,\rm (E,B)} \rightarrow \varphi^{(\rm ph)} \gamma $ described in the lowest order by the part

\begin{equation}
-\sqrt{f} \varphi^{(\rm ph)} \left(\vec{E}\cdot \vec{A}^{\rm (E)} - \vec{B}\cdot \vec{A}^{\rm (B)}\right)
\end{equation}

\ni of the interaction (21), where $\vec{E} = -\partial_0 \vec{A} - \vec{\partial}A_0$ and $\vec{B} = \vec{\partial}\times \vec{A}$. Then, the corresponding $S$ matrix elements (in an obvious notation) are

\begin{equation} 
S^{\rm (E)}_{f i} = \sqrt{f}\left[\frac{1}{(2\pi)^9}\frac{1}{8\omega\omega_\varphi \omega_A} \right]^{1/2} \frac{1}{i}
\left[\left(\omega \vec{e} - \vec{k} e_0\right)\cdot\vec{e}^{\,\rm (E)}\right](2\pi)^4\delta^4(k+k_\varphi - k_A)
\end{equation}

\ni and

\begin{equation}
S^{\rm (B)}_{f i} = \sqrt{f}\left[\frac{1}{(2\pi)^9}\frac{1}{8\omega\omega_\varphi \omega_A}\right]^{1/2} \frac{1}{i}\left[\left(\vec{k}\times\vec{e}\right)\cdot\vec{e}^{\,\rm (B)}\right](2\pi)^4\delta^4(k+k_\varphi - k_A)\;.
\end{equation}

Hence, the formulae for differential decay rates 

\begin{equation} 
\frac{d^6\Gamma^{\rm (E,B)}}{d^3\vec{k}\,d^3\vec{k_\varphi}} = (2\pi)^3\sum_e \frac{1}{3}\sum_{e^{\rm (E,B)}} \frac{{|S^{\rm (E,B)}_{f i}}|^2}{(2\pi)^4 \delta^4(0)}  
\end{equation}

\ni give two expressions that turn out to be equal:

\begin{eqnarray} 
\frac{d^6\Gamma^{\rm (E)}}{d^3\vec{k}\, d^3\vec{k_\varphi}} & = & \frac{f}{(2\pi)^2}\frac{1}{24\omega\omega_\varphi \omega_A} \sum_e \sum_{{e}^{\,\rm (E)}}\left(\omega \vec{e}\cdot\vec{e}^{\,\rm (E)}\right)^{\!\!2} \delta^4(k + k_\varphi - k_A) \nonumber \\ 
& = & \frac{f}{(2\pi)^2} \frac{\omega}{12\omega_\varphi \omega_A} \delta^4(k + k_\varphi - k_A) 
\end{eqnarray}

\ni and 

\begin{eqnarray} 
\frac{d^6\Gamma^{\rm (B)}}{d^3\vec{k}\, d^3\vec{k_\varphi}} & = & \frac{f}{(2\pi)^2}\frac{1}{24\omega\omega_\varphi \omega_A} \sum_e \sum_{{e}^{\,\rm (B)}}\left[(\vec{k}\times \vec{e})\cdot\vec{e}^{\,\rm (B)}\right]^{\!\!2} \delta^4(k + k_\varphi - k_A) \nonumber \\ & = & \frac{f}{(2\pi)^2} \frac{\omega}{12\omega_\varphi \omega_A} \delta^4(k + k_\varphi - k_A) \,.
\end{eqnarray}

\ni Here, in addition to the relations (26) for $A$-boson polarizations, we use the standard formulae 

\begin{equation} 
\vec{e}_a \cdot \vec{e}_b = \delta_{a b} \;\;(a,b = 1,2)\;\;\;,\;\;\; \sum_{a=1}^2 e_{a k} e_{a l} = \delta_{k l} - \frac{k_k k_l}{\vec{k}^2}
\end{equation}

\ni for two orthonormal linear photon polarizations $ \vec{e} = \vec{e}_a\;(a=1,2)$, where $e =(e^0,\vec{e})$ with $e^0 = 0$.

Thus, we obtain from Eqs. (31) and (32) two equal total decay rates that at rest ($\vec{k}_A = 0$) are

\begin{equation}  
\Gamma^{\rm (E,B)} = \int d^3\vec{k}\,d^3\vec{k}_\varphi \frac{d^6\Gamma^{\rm (E,B)}}{d^3\vec{k}\,d^3\vec{k}_\varphi} = \frac{f}{12\pi} \frac{\omega^3}{M^2} = \frac{f}{96\pi} M \left( 1 - \frac{m_\varphi^2}{M^2} \right)^{\!\!3} \,,
\end{equation}

\ni since $\vec{k} + \vec{k}_\varphi = \vec{k}_A = 0$ and $\omega + \omega_\varphi = \omega_A = M$ (so, $\omega =(M/2)(1- m^2_\varphi/M^2)$, $\omega_\varphi =(M/2)(1+ m^2_\varphi/M^2)$ and $|\vec{k}| = \omega \;,\; |k_\varphi| = \sqrt{\omega^2_\varphi - m^2_\varphi}$ ).

\vspace{0.4cm}

\ni {\bf 4. Decays of $A^{\rm (E,B)} \rightarrow e^+ e^-$ (with $M>2m_e$)} 

\vspace{0.4cm}

Now, consider two processes $A^{\rm (E,B)} \rightarrow \gamma^* \rightarrow  e^+ e^- $ described in the lowest order by the part 

\begin{equation}
\sqrt{f}\,<\!\!\varphi\!\!>_{\rm vac} \left(\vec{E}\cdot \vec{A}^{\rm (E)} - \vec{B}\cdot \vec{A}^{\rm (B)}\right) 
\end{equation}

\ni of the interaction (21), where again $\vec{E} = -\partial_0 \vec{A} - \vec{\partial}A_0$ and $\vec{B} = \vec{\partial}\times \vec{A}$, jointly with the Standard-Model electromagnetic interaction of electrons $e\bar{\psi}_e \gamma^\mu\psi_e A_\mu$. In this case, the corresponding $S$ matrix elements (in an obvious notation) take the form 

\begin{equation}
S^{\rm (E)}_{f i} \!\!=\!\! e\!\sqrt{\!f} \!<\!\!\varphi\!\!>_{\rm vac}\!\!\left[\frac{1}{(2\pi)^9}\frac{m^2_e}{E_1 E_2 2\omega_{\!A}}\!\right]^{\!1/2}\!\!\left[\!\bar{u}(p_1\!)\frac{1}{i}\left(\!\omega_{\!A}\vec{\gamma} \!-\! \vec{k}_{\!A}\beta\!\right)\!\!\cdot\!\vec{e}^{\,\rm (E)}v(p_2\!)\!\right]\!\frac{1}{k^2_A}\! (2\pi)^4 \delta^4(p_1 \!+\!p_2 \!-\! k_A)
\end{equation}

\ni and

\begin{equation}
S^{\rm (B)}_{f i} \!\!=\!\! e\sqrt{\!f}\! <\!\!\varphi\!\!>_{\rm vac}\!\!\left[\frac{1}{(2\pi)^9}\frac{m^2_e}{E_1 E_2 2\omega_{\!A}}\right]^{1/2}\!\! \left[\!\bar{u}(p_1\!)\frac{1}{i}\left(\vec{k}_A\times\!\vec{\gamma}\right)\!\cdot\!\vec{e}^{\,\rm (B)}v(p_2\!)\! \right]\!\frac{1}{k^2_A} (2\pi)^4\delta^4(p_1 \!+\! p_2 \!-\! k_A)\,.
\end{equation}

Hence, the formulae for differential decay rates

\begin{equation}
\frac{d^{\,6}\Gamma^{\rm (E,B)}}{d^3\vec{p}_1 d^3\vec{p}_2} = (2\pi)^3\sum_{u,\,v}\frac{1}{3} \sum_{e^{\rm (E,B)}} \frac{|S^{\rm (E,B)}_{f i}|^2}{(2\pi)^4 \delta^4(0)} 
\end{equation}

\ni lead to two different expressions:

\begin{equation}
\frac{d^{\,6}\Gamma^{\rm (E)}}{d^3\vec{p}_1 d^3\vec{p}_2} = \frac{e^2 f<\!\!\varphi\!\!>^2_{\rm vac}}{(2\pi)^2}
 \frac{1}{6E_1 E_{2\,}\omega_{\!A}} \left[1 - 2\frac{E_1 E_2}{M^2} + 2\left(\frac{m_e\omega_A}{M^2}\right)^{\!\!2}\right] \delta^4(p_1 + p_2 - k_A) 
\end{equation}

\ni and

\begin{equation}
\frac{d^{\,6}\Gamma^{\rm (B)}}{d^3\vec{p}_1 d^3\vec{p}_2} =  \frac{e^2 f<\!\!\varphi\!\!>^2_{\rm vac}}{(2\pi)^2}
 \frac{1}{6E_1 E_{2\,}\omega_{\!A}} \left[\frac{\vec{k}^2_A}{M^2} - \left(\frac{\vec{p}_1\times \vec{k}_A}{M^2}\right)^{\!\!2} - \left(\frac{\vec{p}_2\times \vec{k}_A}{M^2}\right)^{\!\!2} \right] \delta^4(p_1 + p_2 - k_A) \,.
\end{equation}

\ni Here, we apply the standard formulae

\begin{equation}
\bar{u}(p_1) u(p_1) \!=\! 1, \sum_u u(p_1) \bar{u}(p_1) \!=\! \frac{\gamma\!\cdot\! p_1 \!+\! m_e}{2m_e} , \bar{v}(p_2) v(p_2) \!=\! -1 \,, \sum_v v(p_2) \bar{v}(p_2) \!=\! \frac{\gamma\!\cdot\! p_2 \!-\! m_e}{2m_e}     
\end{equation}

\ni for unit Dirac bispinors $u(p_1)$ and $v(p_2)$.

At rest ($\vec{k}_A =0$), the total decay rate $\Gamma^{(E)}$ following from Eq. (39) takes the form

\begin{equation}
\Gamma^{\rm (E)} = \int d^3\vec{p}_{\!1}\,d^3\vec{p}_2 \frac{d^{\,6}\Gamma^{\rm (E)}}{d^3\vec{p}_{\!1}\,d^3\vec{p}_2} = \frac{e^2 f <\!\!\varphi\!\!>^2_{\rm vac}}{24\pi M}\left[1 + \left(\frac{2m_e}{M}\right)^{\!\!2}\right] \left[1 - \left(\frac{2m_e}{M}\right)^{\!\!2}\right]^{\!1/2} \!, 
\end{equation}

\ni since $\vec{p}_1 + \vec{p}_2 = \vec{k}_A = 0$ and $E_1 + E_2 = \omega_A = M$ (so, $E_1 = E_2 =M/2$ and $ |\vec{p}_1| = |\vec{p}_2| = \sqrt{(M/2)^2 - m^2_e}$). 

Note that at rest ($\vec{k}_A =0$) the differential decay rate (40) for $A^{(B)}$ bosons vanishes. Thus, also the total decay rate $\Gamma^{\rm (B)}$ is zero at rest ($\vec{k}_A =0$).  This vanishing of decay rates for $A^{\rm(B)}$ bosons at rest ($\vec{k}_A =0$) is a consequence of virtual equality $ k_A = k = p_1 + p_2$ valid in the process $A \rightarrow \gamma^* \rightarrow e^+ e^-$, where the combination $(k_\mu g_{\nu \lambda} - k_\nu g_{\mu \lambda})/k^2$ (following from the contraction of $F_{\mu\,\nu} = \partial_\mu A_\nu - \partial_\nu A_\mu$ with $A_\lambda$ appearing in $e\bar{\psi}_e \gamma^\lambda \psi_e A_\lambda$) describes the propagation of $\gamma^*$ on the $A$-boson mass shell $k^2 = k^2_A = M^2$ with $\vec{k} = \vec{k}_A = 0$ and $\omega = \omega_A = M$ (in this combination, $\mu = k = 1,2,3$ and $\nu = l = 1,2,3$ for $A^{\rm (B)}$ bosons).  
 
\vspace{0.4cm}

\ni {\bf 5. Final remarks} 

\vspace{0.4cm}

Our model --- involving beside the sterino weak coupling $-(1/2)\sqrt{f}\,\zeta \bar{\psi} \sigma_{\mu \nu}\psi A^{\mu \nu}$ the new weak interaction $-(1/2)\sqrt{f}\,\varphi F_{\mu \nu} A^{\mu \nu}$ of photons paired with sterons --- implies the existence in Nature of a non-gauge antisymmetric-tensor field $A_{\mu \nu}$ (of dimension one) mediating interactions both within the hidden sector as well as between this sector and photons.  Of course, photons interact also electromagnetically with the Standard-Model sector according to the conventional gauge scheme. In contrast, our new photon interaction has a non-typical form that, after eliminating the massive mediator $A_{\mu \nu}$, becomes asympthotically a bosonic Fermi-type coupling $-(1/4)(f/M^2)(\varphi F_{\mu \nu})(\varphi F^{\mu \nu})$ of two steron-photon pairs. Note that, due to this coupling, a phenomenon of finite additional renormalization ("primordial renormalization") [1] corresponding to $\varphi = <\!\!\varphi\!\!>_{\rm vac}$ appears in our model. 

The possibility of coupling constant $f$ guessed to assume the value $e^2 = 4\pi \alpha$ could be exciting [1]. Then, $f/M^2$ equal to $e^2/M^2 = 4\pi \alpha/M^2$ with a large mass scale $M$ would play the role of bosonic Fermi-type constant for steron-photon pairs. In this case, Eq. (10) for the sterino magnetic moment spontaneously generated by  $ <\!\!\varphi\!\!>_{\rm vac} \neq 0$ would give $m_\psi = [e^2\zeta/(2M^2)]<\!\!\varphi\!\!>_{\rm vac}$ (where $\zeta$ might be 1).

In this paper, we have not considered the attractive option, where more than one generation of sterinos and sterons are present in the hidden sector. In particular, the existence of three generations of sterinos might be natural [5] according to an intrinsic  approach to fermion generations invented some years ago. Such multiple sterinos could provide a model of the so-called exciting cold dark matter [3]. In this case, the new possibility of moderately light $A$ bosons might turn out to be advantageous for interpretation [3] of indirect ({\it e.g.} PAMELA and INTEGRAL) and direct ({\it e.g.} DAMA) detection experiments for the cold dark matter.

We have assumed in this paper that there is no direct coupling of sterons with Standard-Model Higgs bosons which may lead to their mixing and so can be responsible for the "Higgs portal"~to the hidden sector [3]. In our model, sterons and Higgs bosons are not directly related, though both kinds of these scalars participate in the spontaneous mass generation (for hidden and Standard-Model particles, respectively).
 
\vfill\eject

\vspace{0.4cm}

{\centerline{\bf References}}

\vspace{0.4cm}

\baselineskip 0.73cm

{\everypar={\hangindent=0.65truecm}
\parindent=0pt\frenchspacing

{\everypar={\hangindent=0.65truecm}
\parindent=0pt\frenchspacing

~[1]~W.~Kr\'{o}likowski, {\it Acta Phys. Polon.} {\bf B 39}, 1881 (2008) (arXiv: 0712.0505 [{\tt hep--ph}]); arXiv: 0803.2977v2 [{\tt hep--ph}]; arXiv: 0806.2036v2 [{\tt hep--ph}]; {\it Acta Phys. Polon.} {\bf B 40}, 111 (2009) (arXiv: 0809.1931v2 [{\tt hep--ph}]) .

\vspace{0.2cm}

~[2]~For some reviews {\it cf.} E.W.~Kolb and S.~Turner, {\it Early Universe}, Addison-Wesley, Reading, Mass., 1994; K.~Griest and D.~Seckel, {\it Phys. Rev.} {\bf D 43}, 3191 (1991); G.~Bartone, D.~Hooper and J.~Silk, {\it Phys. Rep.} {\bf 405}, 279 (2005); M.~Taoso, G.~Bartone and A.~Masiero, arXiv: 0711.4996 [{\tt astro-ph}]; J.L.~Feng, H.~Tu and H.-B.~Yu, arXiv: 0808.2318 [{\tt hep-ph}].  

\vspace{0.2cm}

~[3]~For some recent publications {\it cf.} D.P.~Finkbeiner and N.Weiner, {\it Phys. Rev.} {\bf D 76}, 083519 (2007); J. March-Russell, S.M. West, D. Cumberbath and D.~Hooper, {\it J. High Energy Phys.} {\bf 0807}, 058 (2008); D.~Hooper and K.M. Zurek,  {\it Phys. Rev.} {\bf D 77}, 087302 (2008); J.~McDonald and N.~Sahu, {\it JCAP} {\bf 0806}, 026 (2008); Y.G.~Kim, K.Y.~Lee and S.~Shin, {\it J. High Energy Phys.} {\bf 0805}, 100 (2008); J.L.~Feng and J.~Kumar, arXiv: 0803.4196 [{\tt hep-ph}]; R.~Foot, arXiv: 0804.4518 [{\tt hep-ph}]; M.~Cirelli, M.~Kadastik, M.~Raidal and A.~Strumia, arXiv: 0809.2409 [{\tt hep-ph}]; N.~Arkani-Hamed, D.P.~Finkbeiner, T.R.~Slatyer and N.~Weiner, {\it Phys. Rev.} {\bf D 79}, 015014 (2009); D.P.~Finkbeiner, T.R.~Slatyer, N.~Weiner and I.~Yavin,  arXiv: 0903.1037 [{\tt hep-ph}]; M.~Ibe, 
Y.~Nakayama, H.~Murayama and T.T.~Yanagida,  arXiv: 0902.2914 [{\tt hep-ph}]; and references therein.

\vspace{0.2cm}

~[4]~C.~Amsler {\it et al} (Particle Data Group), {\it Review of Particle Physics,  Phys. Lett.}, {\bf B 667}, 1 (2008).

\vspace{0.2cm}

~[5]~W.~Kr\'{o}likowski, arXiv: 0811.3844 [{\tt hep--ph}]; also {\it Acta Phys. Polon.} {\bf B 33}, 2559 (2002); {\tt hep--ph/0504256}; {\tt hep--ph/0604148}; {\it Acta Phys. Polon.} {\bf B 38}, 3133 (2007); and references therein.

\vfill\eject

\end{document}